\documentclass[twocolumn]{openjournal}
\usepackage[english]{babel}     % sets language rules/patterns to English

\usepackage{natbib}     % formats bib and enables citation variety
\usepackage{xcolor}     % color support (HEX, named, RGB/HTML)
\usepackage{textgreek}     % allows greek letters in-text (\textalpha)
\usepackage{amsmath}

\usepackage{hyperref}     % turns refs, citations, URLs into links   
\usepackage{orcidlink}
\hypersetup{
    unicode,     % allows all unicode characters in section bookmarks
    colorlinks=true,     % colors link text instead of making a box
    linkcolor=linkcolor,     % I change following to be colors I want
    citecolor=linkcolor,
    filecolor=linkcolor,
    urlcolor=linkcolor,
}
\usepackage{color,colortbl}     % allows color application to page/tables
\definecolor{linkcolor}{rgb}{0.0,0.3,0.5}     % defines linkcolor above
\DeclareGraphicsExtensions{.bmp,.png,.jpg,.pdf}     % ext latex looks for
\graphicspath{ {./figs/} }     % will look inside '/figs/' directory
\usepackage{multirow}
\usepackage[caption=false]{subfig}
\usepackage{booktabs}
\begin{document}

\title{Constraining the Inner Dark Matter Profiles of Strong-Lensing Galaxy Clusters}

\author{Emma M. Strickland\textsuperscript{1} \orcidlink{0009-0007-0876-7811}, Zelie Goldberg Little\textsuperscript{1}\orcidlink{0009-0000-6279-2013}, Jackson H. O'Donnell\textsuperscript{1,2} \orcidlink{0000-0003-4083-1530}, Tesla E. Jeltema\textsuperscript{1,2} \orcidlink{0000-0001-6089-0365}}

\affiliation{\textsuperscript{1} University of California, Santa Cruz, Santa Cruz, CA 95064, USA}
\affiliation{\textsuperscript{2} Santa Cruz Institute for Particle Physics, Santa Cruz, CA 95064, USA}

% -----------------------------------------------------------------------

\begin{abstract}

We present mass modeling of three strong lensing galaxy clusters found in the Dark Energy Survey utilizing Hubble Space Telescope imaging for lens modeling, and complementary constraints from Chandra X-ray observations. All three clusters are relaxed and have giant arc or ring lens structures ideal for constraints on the dark matter distribution. The strong lens modeling and X-ray temperatures yield consistent total cluster mass estimates. We find these clusters have best-fit X-ray temperatures of $2.7 - 6.2$ keV and total masses of M$_{200} = (1-6) \times 10^{14} M_{\odot}$. Our lens models parameterize dark matter distribution with a generalized Navarro-Frenk-White (GNFW) density profile. While the GNFW profile allows the inner dark matter slope to vary, we find in all cases that the inner slopes are consistent with the standard NFW density profile $\gamma^{DM} = 1$. The X-ray surface brightness distributions for all three clusters are generally symmetric with no significant substructure, supporting the conclusion that these are relaxed clusters.  In addition, we find generally low values of the external shear magnitude with median values ranging $|\gamma^{ext}| = 0.04-0.09$ and consistent with zero for two of the clusters.

\end{abstract}

\maketitle

% -----------------------------------------------------------------------

\section{Introduction}
\label{sec:intro}

Strong gravitational lensing by galaxy clusters provides a powerful tool for studying the formation and evolution of large-scale structures, and strong lensing clusters are also powerful gravitational telescopes, allowing the study of the high-redshift universe. In particular, strong lensing offers a direct probe of the total mass distribution in the innermost region of the cluster. This allows the central mass distribution of the cluster to be well constrained, particularly when extended features such as giant arcs or partial Einstein rings are present \citep[see e.g.][for a recent review]{Natarajan24}.

Strong lensing is thus a powerful tool, not only for cluster mass estimation, but also for testing the shape of the inner dark matter density profile where varying dark matter models make divergent predictions. In particular, cold dark matter (CDM) models predict cuspy inner density profiles while non-CDM models, including self-interacting dark matter (SIDM), warm dark matter (WDM), and fuzzy dark matter (FDM), generically predict shallower, cored profiles \citep[e.g.][]{Spergel00,Hu00,Schive14,Kaplinghat16,Hui17,Tulin18,Despali26}. Because clusters are strongly dark matter dominated, and their strong lensing signal is generated on scales comparable to the core radii where these models diverge, cluster-scale strong lensing has emerged as a sensitive test for underlying dark matter properties \citep{Newman13b,Sagunski21,Andrade22,ODonnell_2026}.

Extracting robust constraints on the inner dark matter density profile from lensing alone is complicated by the need to separate the dark matter distribution from the baryonic mass contribution and by uncertainties in the cluster structure, e.g.~substructure in merging clusters. Combining strong lens modeling with X-ray observations mitigates both issues. The X-ray surface brightness morphology serves as an established diagnostic of dynamical relaxation \citep[e.g.][]{Mohr93,Buote96,Jeltema05,Poole06,Yuan22,Veliz25}. A lack of substructure in X-ray with symmetric, centrally peaked emission indicates an undisturbed system, while the X-ray temperature provides a mass estimate that is independent of lensing, offering a valuable consistency check on the total mass \citep[e.g.][]{Sarazin86,Rosati02,Voit05,Allen11}.

In this paper, we present strong lens models for three galaxy clusters identified in the Dark Energy Survey (DES), selected for their giant arc or ring-like lensing features that provide strong geometric constraints and indicate relaxed systems based on the symmetry of their arc systems. We combine Hubble Space Telescope (HST) imaging, for lens modeling, with Chandra X-ray observations, which we use to confirm the dynamical state and to obtain an independent temperature-based mass estimate. We model the dark matter distribution in each cluster with a generalized NFW (GNFW) profile that allows the inner slope to vary as a free parameter. By comparing the recovered inner slopes to the standard CDM prediction, we test whether these relaxed cluster cores favor the cuspy profiles expected under collisionless CDM or show evidence for the shallower cores predicted by non-CDM dark matter models.

In Section \ref{sec:data}, we outline the properties of the strong lens clusters chosen and the optical and X-ray data sets used. Section \ref{sec:method} discusses the strong lens modeling and the model components, as well as the X-ray spectral extraction and modeling. In Section \ref{sec:result}, we give our results, and Sections \ref{sec:disc} and \ref{sec:conc} present our discussion and conclusions.

% -----------------------------------------------------------------------

\section{Data Selection \& System Characteristics}
\label{sec:data}

\subsection{Description of Cluster Systems}
\label{subsec:clusters}

\begingroup % following affects only this particular table
    \renewcommand{\arraystretch}{1.5} % adjusts row height
    \setlength\extrarowheight{2pt}
    \setlength{\tabcolsep}{8pt} 
    \begin{table*}[t]
        \centering
        \begin{tabular}{c || c c c c c} 
            System & RA (deg) & Dec (deg) & $z_{lens}$ & $z_{source}$ & References\\
            \hline     
            DESJ0424-3317 & 66.161 & -33.295 & 0.564 & 1.188 & (D) \\
            DESJ2349-5113 & 357.375 & -51.227 & 0.418 & 1.390 & (C), (D), (E) \\
            DESJ2351-5452 & 357.908 & -54.882 & 0.384 & 0.906 & (A), (B), (C) \\
        \end{tabular}
        \caption{A summary of the physical attributes of the lensing systems used in this work. This includes the sky position of the BCG of each system, the redshifts of the lens ($z_{lens}$) and source ($z_{source}$) planes, and references for the redshift values. \\
        \textbf{References.} (A) \cite{Menanteau2010}, (B) \cite{Buckley-Geer2011}, (C) \cite{Diehl2017}, (D) \cite{Jacobs2019}, (E) \cite{Rojas2022}.}
        \label{table:sys}
    \end{table*}
\endgroup

\begin{figure*}[t!]
    \centering
    \subfloat[DESJ0424-3317]{%
        \includegraphics[width=0.7\textwidth]{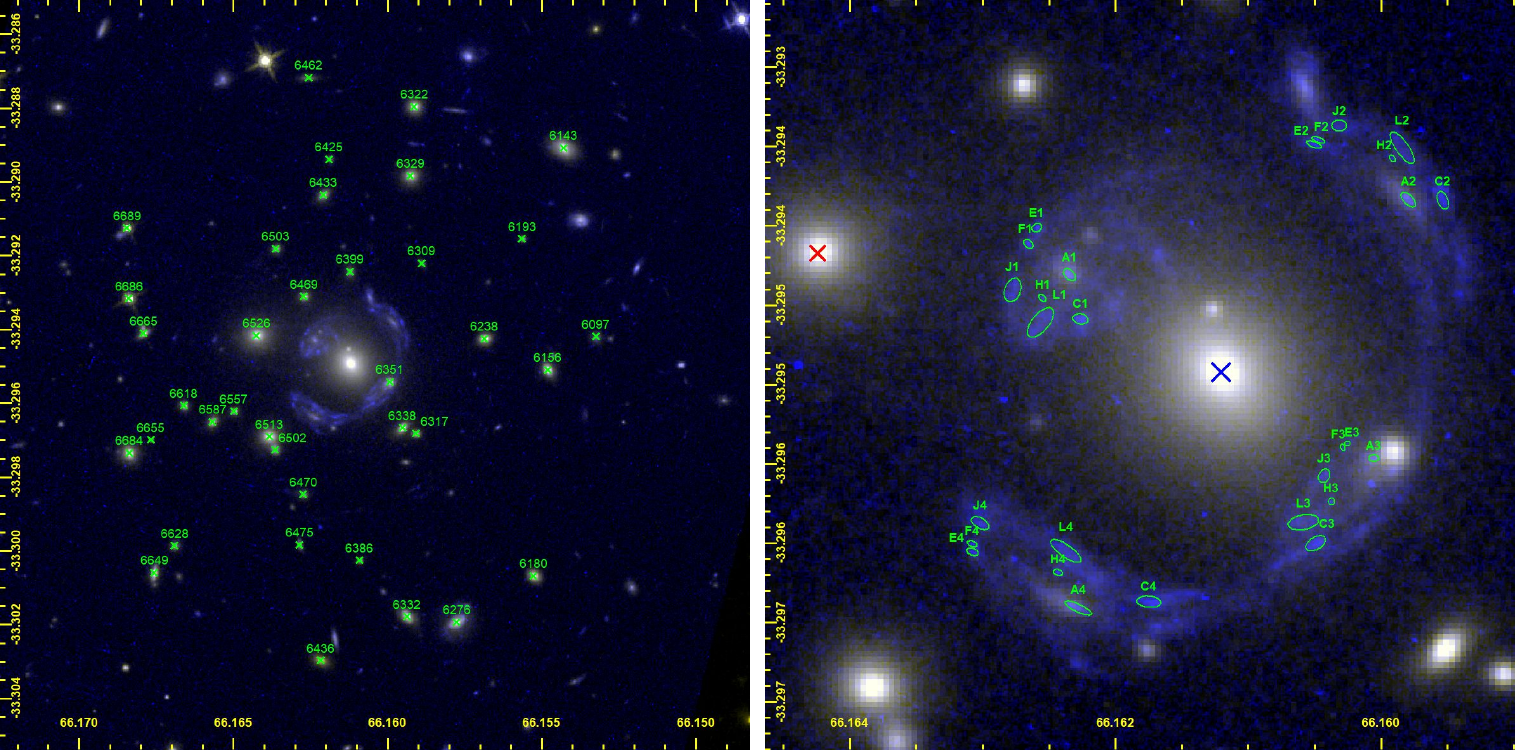}%
    }\vspace{0.53cm}
    \subfloat[DESJ2349-5113]{%
        \includegraphics[width=0.7\textwidth]{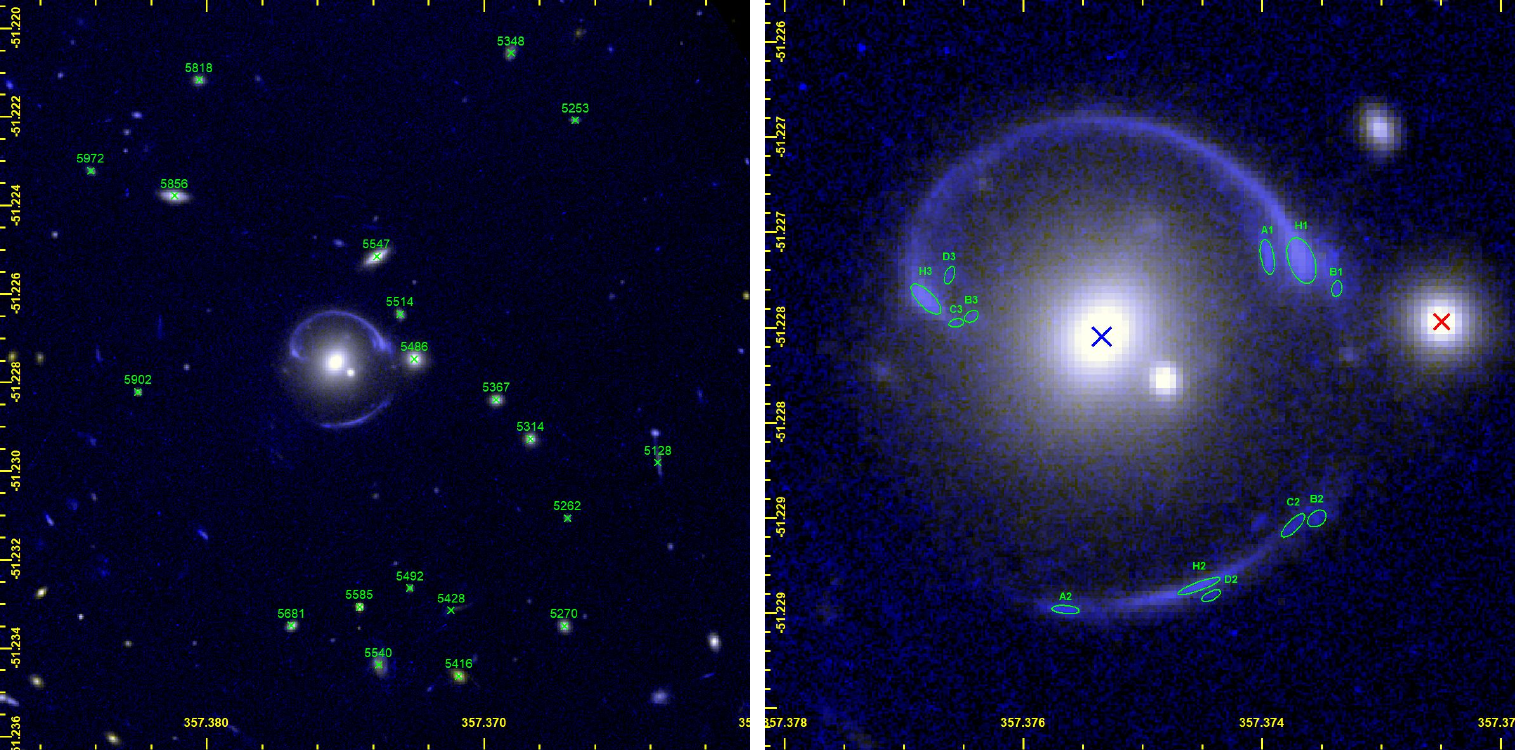}%
    }\vspace{0.53cm}
    \subfloat[DESJ2351-5452]{%
        \includegraphics[width=0.7\textwidth]{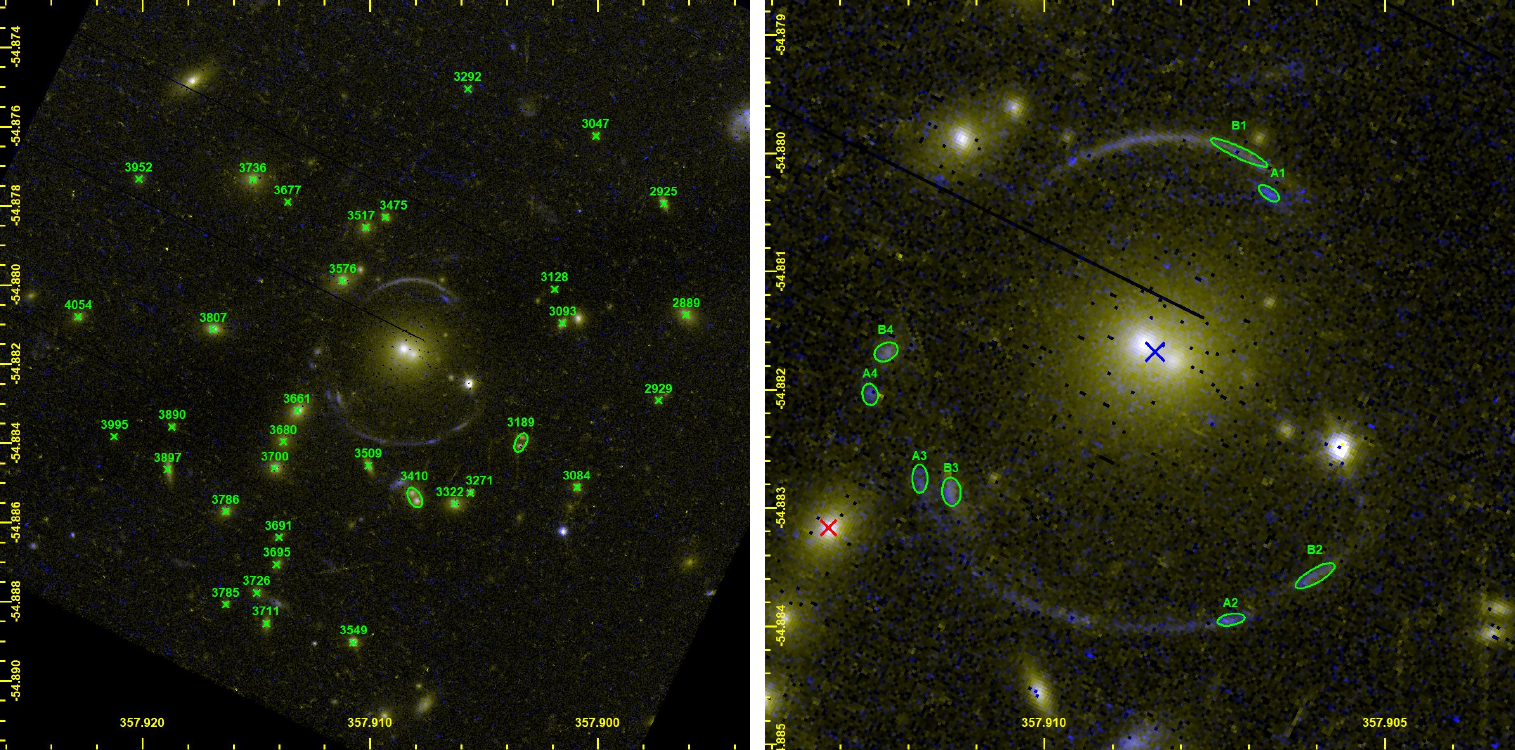}%
    }\vspace{0.53cm}
    \caption{HST imaging of each cluster system with grid axes in degrees. The right-hand panels show the features used to model each cluster in green, along with our chosen pivot galaxy in red, and the BCG in blue. The left-hand panels show the cluster member galaxies obtained from the DR10 catalogs. The selection of these objects are described in Section \ref{sec:method} and further physical details on the used features are available in Table \ref{tab:features}.}
    \label{fig:systems}
\end{figure*}

\begingroup
\setlength{\tabcolsep}{8pt} 
\begin{table*}[h]
    \centering
    \begin{tabular}{c c c c c c}
        \shortstack{Feature\\Label} & RA (deg) & Dec (deg) & \shortstack{Semi-Major\\Axis} & \shortstack{Semi-Minor\\Axis} & \shortstack{Rotation\\Angle} \\
        \toprule \toprule
        DESJ0424-3317 & & & & & \\
        \midrule[\heavyrulewidth]
        A1 & 66.16235 & -33.29431 & 0.10" & 0.17" & $45^\circ$ \\
        A2 & 66.15980 & -33.29384 & 0.21" & 0.11" & $315^\circ$ \\
        A3 & 66.16006 & -33.29547 & 0.12" & 0.09" & $0^\circ$ \\
        A4 & 66.16228 & -33.29642 & 0.33" & 0.09" & $155^\circ$ \\
        \midrule
        C1 & 66.16226 & -33.29459 & 0.17" & 0.12" & $350^\circ$ \\
        C2 & 66.15954 & -33.29385 & 0.11" & 0.21" & $20^\circ$ \\
        C3 & 66.16049 & -33.29601 & 0.14" & 0.25" & $300^\circ$ \\
        C4 & 66.16175 & -33.29638 & 0.28" & 0.13" & $355^\circ$ \\
        \midrule
        E1 & 66.16259 & -33.29402 & 0.09" & 0.11" & $120^\circ$ \\
        E2 & 66.16050 & -33.29349 & 0.18" & 0.06" & $340^\circ$ \\
        E3 & 66.16026 & -33.29538 & 0.07" & 0.06" & $0^\circ$ \\
        E4 & 66.16308 & -33.29606 & 0.13" & 0.08" & $340^\circ$ \\
        \midrule
        F1 & 66.16266 & -33.29412 & 0.08" & 0.12" & $45^\circ$ \\
        F2 & 66.16047 & -33.29346 & 0.16" & 0.05" & $340^\circ$ \\
        F3 & 66.16029 & -33.29540 & 0.05" & 0.08" & $20^\circ$ \\
        F4 & 66.16308 & -33.29601 & 0.11" & 0.06" & $340^\circ$ \\
        \midrule
        H1 & 66.16255 & -33.29446 & 0.06" & 0.10" & $45^\circ$ \\
        H2 & 66.15992 & -33.29358 & 0.05" & 0.09" & $45^\circ$ \\
        H3 & 66.16037 & -33.29574 & 0.07" & 0.08" & $330^\circ$ \\
        H4 & 66.16243 & -33.29619 & 0.11" & 0.07" & $340^\circ$ \\
        \midrule
        J1 & 66.16277 & -33.29441 & 0.18" & 0.28" & $340^\circ$ \\
        J2 & 66.16032 & -33.29338 & 0.17" & 0.13" & $0^\circ$ \\
        J3 & 66.16043 & -33.29558 & 0.12" & 0.16" & $150^\circ$ \\
        J4 & 66.16302 & -33.29588 & 0.22" & 0.12" & $330^\circ$ \\
        \midrule
        L1 & 66.16256 & -33.29461 & 0.19" & 0.41" & $320^\circ$ \\
        L4 & 66.16237 & -33.29606 & 0.41" & 0.14" & $325^\circ$ \\
        L2 & 66.15984 & -33.29351 & 0.42" & 0.14" & $125^\circ$ \\
        L3 & 66.16059 & -33.29587 & 0.35" & 0.17" & $10^\circ$ \\
        \midrule[\heavyrulewidth]
        DESJ2349-5113 & & & & & \\
        \midrule[\heavyrulewidth]
        A1 & 357.37395 & -51.22713 & 0.12" & 0.34" & $10^\circ$ \\
        A2 & 357.37564 & -51.22898 & 0.26" & 0.08" & $355^\circ$ \\
        \midrule
        B1 & 357.37337 & -51.22730 & 0.09" & 0.15" & $350^\circ$ \\
        B2 & 357.37354 & -51.22851 & 0.19" & 0.14" & $35^\circ$ \\
        B3 & 357.37643 & -51.22744 & 0.14" & 0.10" & $35^\circ$ \\
        \midrule
        C2 & 357.37374 & -51.22854 & 0.30" & 0.11" & $45^\circ$ \\
        C3 & 357.37656 & -51.22748 & 0.14" & 0.08" & $10^\circ$ \\
        \midrule
        D2 & 357.37442 & -51.22891 & 0.19" & 0.08" & $25^\circ$ \\
        D3 & 357.37661 & -51.22723 & 0.09" & 0.17" & $340^\circ$ \\
        \midrule
        H1 & 357.37367 & -51.22715 & 0.24" & 0.45" & $20^\circ$ \\
        H2 & 357.37452 & -51.22886 & 0.42" & 0.08" & $20^\circ$ \\
        H3 & 357.37681 & -51.22735 & 0.37" & 0.15" & $135^\circ$ \\
        \midrule[\heavyrulewidth]
        DESJ2351-5452 & & & & & \\
        \midrule[\heavyrulewidth]
        A1 & 357.90670 & -54.88034 & 0.17" & 0.362" & $55^\circ$ \\
        A2 & 357.90726 & -54.88395 & 0.17" & 0.418" & $100^\circ$ \\
        A3 & 357.91183 & -54.88276 & 0.43" & 0.231" & $90^\circ$ \\
        A4 & 357.91256 & -54.88204 & 0.32" & 0.236" & $100^\circ$ \\
        \midrule
        B1 & 357.90714 & -54.88000 & 0.17" & 0.946" & $65^\circ$ \\
        B2 & 357.90602 & -54.88358 & 0.67" & 0.210" & $30^\circ$ \\ 
        B3 & 357.91136 & -54.88287 & 0.43" & 0.284" & $95^\circ$ \\
        B4 & 357.91232 & -54.88168 & 0.37" & 0.263" & $25^\circ$ \\
        \bottomrule        
    \end{tabular}
    \caption{Tabled details of the source features used in the modeling of each system. This includes the sky position in degrees, as well as the length of the axes and position angle of the elliptical regions for each feature used.}
    \label{tab:features}
\end{table*}
\endgroup

Lens modeling and X-ray analysis is presented for three strong-lensing cluster systems: DESJ0424-3317, DESJ2349-5113, and DESJ2351-5452  (hereafter referred to as DESJ0424, DESJ2349, and DESJ2351) selected from cluster lenses in the Dark Energy Survey (DES) by \cite{ODonnell2022}. Basic properties of these systems, including source and lens redshifts, are listed in Table \ref{table:sys}. We have chosen these specific systems for our modeling for several reasons, including their similar redshifts that range approximately $z=0.4-0.6$, and for the availability of HST imaging data for all systems, providing a high resolution data set for lens modeling. The lensed source galaxies range in redshift from $z=0.9-1.4$.

These systems were selected as likely relaxed clusters displaying symmetric lensing features and a dominant central galaxy, leading to a comparably simplified lens modeling. This is confirmed by our X-ray observations; they exhibit relatively round X-ray emission lacking significant substructure (see Section~\ref{sec:x-ray_temps}, Figure~\ref{fig:x-ray}), likewise indicating that they are relaxed clusters. 

Additionally, each of these systems displays giant arcs and counter arcs. As seen in Figure~\ref{fig:systems}, all three systems have near ring-like lens structures. This figure shows the HST imaging with both a zoom in on the lensing region (right panels) and the larger cluster field (left panels).

\subsection{DESI Legacy Survey Imaging}
\label{subsec:desi}

We use the object catalogs for our cluster regions from Data Release 10 (DR10) of the DESI Legacy Imaging Surveys (hereafter referred to as the Legacy Surveys) \citep{Dey2019}. Sky coordinates and magnitudes of each cluster's member galaxies (shown in Figure \ref{fig:systems}) are obtained from the DR10 object catalogs. Specifically, we used the Legacy Surveys r- and i-band magnitudes to estimate cluster membership and to model the member galaxy mass contributions through the Faber-Jackson relation, as explained in further detail in Sections \ref{sec:memgals} and \ref{subsec:components}.

\subsection{Hubble Space Telescope Imaging}
\label{subsec:hst}

Additionally, this analysis relies on high-resolution imaging data provided by HST to extract lensed source positions. Our data covering systems DESJ0424 and DESJ2349 utilize both the IR and UVIS channels on the telescope's Wide Field Camera 3 (WFC3) with the F140W wide-band filter and F200LP long-pass filter, respectively. 
The data for the aforementioned systems are available under program number 16773 (PI Karl Glazebrook). The data covering the DESJ2351 system use the now retired Wide-Field Planetary Camera 2 (WFPC2) with the wide-band filters F450W and F814W \citep{WFPC2}. This data set is available under program number 11974 (PI Sahar Allam). While the wide-band filters onboard HST provide a wide coverage of wavelengths ranging from blue to near-IR, the long-pass filters provide the necessary depth on faint sources.
These data have a pixel scale of $0.05-0.1$ arcsec, giving the resolution necessary to precisely locate features in the source galaxies and enabling strong-lens modeling.

\subsection{Chandra X-ray Imaging}
\label{subsec:xray_imaging}

The three target clusters were observed by Chandra with the Advanced CCD Imaging Spectrometer (ACIS), specifically using the ACIS-I CCD array in FAINT mode. Observations were made for DESJ2351 in October 2023 for 15.57 ks (ObsID 26925); DESJ2349 was observed for 29.70 ks in September 2024 (ObsID 26924), and three combined observations were completed for DESJ0424 in November 2023 and March 2024 for a total of 54.56 ks (ObsIDs 26926, 27193, and 29343). The collected data were reduced using the Chandra Interactive Analysis for Observations (\texttt{CIAO}) 4.17 software package \citep{CIAO, Fruscione2026}. 

This reduction begins by first reprocessing the level-1 event files into level-2 event files using the \texttt{chandra\_repro} script, automating the data processing procedure recommended in the \texttt{CIAO} analysis. Background flares were then detected and removed with the \texttt{lc\_sigma\_clip} script using the light curve extracted in the $0.5 - 5.0$ keV energy range. Using \texttt{CIAO}'s \texttt{dmstat}, we checked whether the observations exhibited significant variation in focal-plane temperature, and if they did, the data were split into separate temperature ranges for calibration, which was necessary for DESJ2349 and DESJ0424 (ObsIDs 26926 and 27193). 

\begin{figure*}[t!]
    \centering
    \includegraphics[width=1\textwidth]{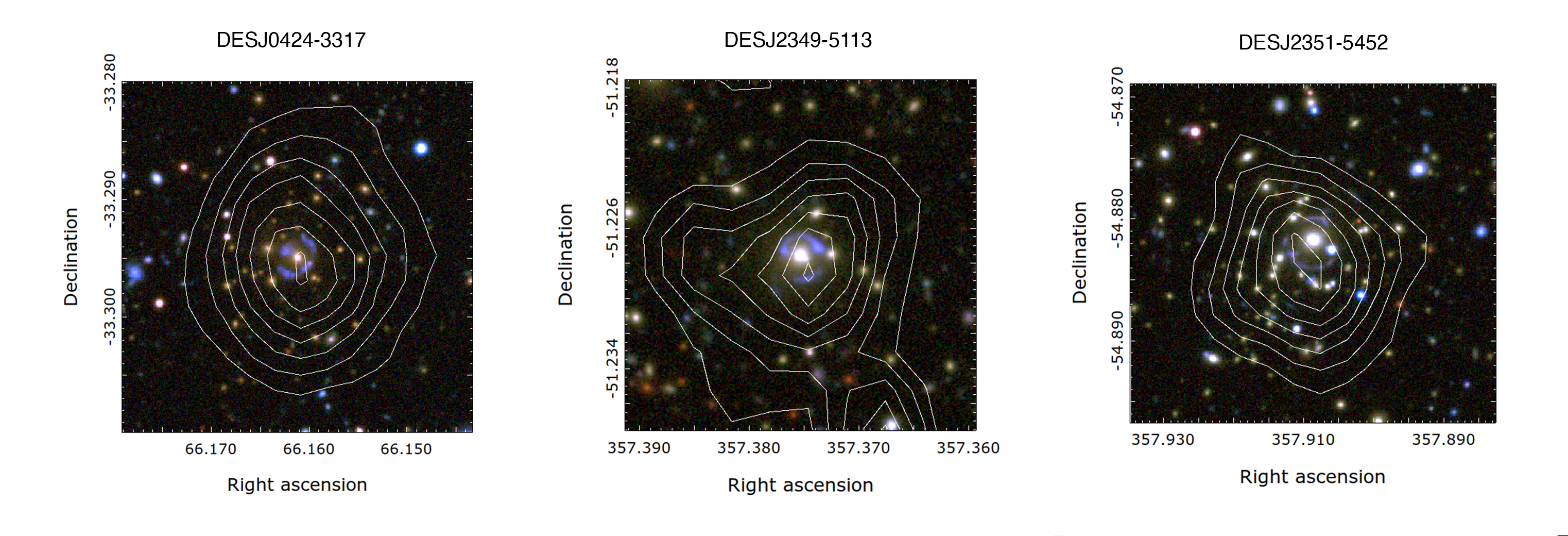}%
    \caption{X-ray contours superimposed on Legacy Sky Surveys images of the three clusters. The X-ray emission does not show signs of substructure, is centered on the clusters, and is roughly spherical. In the case of DESJ2349, an X-ray point source can be seen to the southwest.}
    \label{fig:x-ray}
\end{figure*}

Figure~\ref{fig:x-ray} shows contours of the Chandra X-ray emission overlaid on the Legacy Survey images. 

% -----------------------------------------------------------------------

\section{Methodology}
\label{sec:method}

\subsection{Member Galaxy Selection}
\label{sec:memgals}

% JOD TODO
Our lens modeling procedure can be divided into several stages, beginning first with a series of vetting procedures applied to the data to determine potential cluster member galaxies. We subject all astronomical objects within our catalogs to the following criteria, with objects satisfying these criteria retained as likely member galaxies. Member galaxies are limited by their proximity to the cluster center, specifically to a radial distance within 30 arcsec from the Brightest Cluster Galaxy (BCG), and by a threshold on their color, which we restrict to an r-i color value below -0.5. Furthermore, we apply a threshold on absolute magnitude for potential members, which we have applied differently depending on the system. Objects in systems DESJ0424 and DESJ2349 are restricted to an absolute r-band magnitude less than 24.1, while objects in system DESJ2351 are limited to an absolute r-band magnitude less than 23.7. These various limits ensure our data sets contain only cluster member galaxies and limit the sample to galaxies close enough to the center and large enough to meaningfully impact the lens modeling.

The effect of these member galaxies on lensing is modeled following a scaling procedure defined in Section \ref{subsec:components} that is based on the physical properties of a reference galaxy, which we call the pivot galaxy. To provide a stable reference with which to relate the member galaxies, the most central and brightest galaxy (apart from the BCG) in the cluster is chosen to act as the pivot. This ensures high confidence in the physical properties we use in our scaling relations. The selected members and the pivot galaxy for each system are shown in Figure \ref{fig:systems}.

\subsection{Source Feature Selection}
\label{sec:sources}

Our strong lens models are constrained by the observed positions of multiply-imaged features of each lensed source. These features and their positions are derived from HST imaging and are listed in Table~\ref{tab:features}. We use the modified source-plane likelihood presented and used in \cite{ODonnell_2026} (see Section \ref{subsec:sampl}), which incorporates the elliptical uncertainties of the observed feature positions.

The source-plane positional reconstruction after applying our models is used to judge the performance of our lens models. These lensed features are chosen based on their visual identifiability compared to their surroundings (see Figure \ref{fig:systems}). The selection of lensed features also takes into account each feature's potential to be related to one another by originating from the same source-plane feature, taking into account preliminary lens modeling. Observed lens-plane features are collected into groups of 2 to 4 corresponding to a single source-plane feature (e.g. ``A", ``B"). The basic positional properties of the selected features for each cluster system are presented in Table \ref{tab:features}.

The positions of each source-plane feature are estimated based on the attributes we define for our lens models and the observed positions of the lensed features. An initial prediction of coordinates on the source-plane is made for each lensed feature, from which a single best-fit position is calculated for each grouping, representing the likely position of the original source feature (see Figure \ref{fig:source_posns} in Appendix \ref{ap:ap1}). We expect a well-defined model to trace all individual lensed features back to a common source-plane coordinate, implying the lensing properties of our chosen clusters have been accurately modeled.

It should be noted that the HST imaging for cluster DESJ2351 exhibits an astrometric offset relative to the Legacy Surveys imaging. To correct for this, a manual offset is roughly calculated and applied to align the HST astrometry with the DECam imaging using the \texttt{skimage.transform} package \citep{van_der_Walt_2014}. We provide \texttt{skimage.transform} with two sets of coordinates corresponding to the same selection of prominent cluster members; one set of coordinates has been inaccurately recorded by HST, while the other set is given by the Legacy Surveys. \texttt{skimage.transform} estimates a translation of magnitude $\Delta r = 0.43$ arcsec and a rotation angle of $0.17^\circ$ between the source set and the destination set. We then applied these transformation values to the offset HST source feature coordinates, effectively realigning their positions with the Legacy Survey data.
    
% belongs in sec 3.5 Sampler Config, moved to adjust page proportions
\begingroup
    \setlength\extrarowheight{5pt}
    \begin{table*}
        \centering
        \begin{tabular}{c | c c c c}
         & \shortstack{\texttt{dynesty}\\Default} & DESJ0424-3317 & DESJ2349-5113 & DESJ2351-5452 \\
        \midrule[\heavyrulewidth]
        facc & 0.5 & 0.1 & 0.2 & 0.2 \\
        pfrac & 0.8 & 0.5 & 0.8 & 0.8 \\
        \end{tabular}
        \caption{Table showing both the default values for \texttt{dynesty}'s configuration parameters, \textit{facc} and \textit{pfrac}, and our finalized settings for each lensing system.}
        \label{table: dynesty config}
    \end{table*}
\endgroup

\subsection{Mass Model Components}
\label{subsec:components}

Following the initial data selection and processing, we define a mass model for cluster member galaxies which groups member galaxy contributions together via a scaling relation, excluding the BCG, which is modeled independently. The model for the member galaxies follows a luminosity-dependent scaling law defined by the Faber-Jackson relation: \[L \propto \sigma^{1/\alpha}\] which describes an (elliptical) galaxy's luminosity $L$ scaling with its central stellar velocity dispersion $\sigma$ as $\sigma^{1/\alpha}$ \citep{FJ1976}. We redefine this relation to express the velocity dispersion of each member galaxy in relation to that of the chosen pivot galaxy for each system. A similar scaling relation is also outlined relating the truncation or scale radius of member galaxies to that of the pivot. Both relations are motivated by the scaling relations for the central velocity dispersion and the truncation radius defined by \citet{Bergamini2019}, which are as follows:

\begin{equation}\label{eq:vel-disp}
    \sigma^\text{gal} = \sigma^\text{piv}\left(\frac{L_\text{gal}}{L_\text{piv}}\right)^{\alpha}
\end{equation}
\begin{equation}\label{eq:scale-rad}
    r_s^\text{gal} = r_s^\text{piv}\left(\frac{L_\text{gal}}{L_\text{piv}}\right)^{\beta}
\end{equation}

Here, $\sigma$, $L$ and $r_s$ refer to the velocity dispersion, luminosity, and scale radius of a given member galaxy or pivot galaxy denoted $(\text{gal})$ and $(\text{piv})$, respectively. Additionally, the slope $\beta$ is defined as:
\begin{equation}\label{eq:beta}
    \beta = 1+\gamma-2\alpha
\end{equation}
where we have set $\gamma=0.2$ and take a Gaussian prior on $\alpha$ (with $\mu$ = 0.27 and $\sigma$ = 0.03) based on the results of \citet{Bergamini2019}, who studied the Faber-Jackson relation for three strong lensing clusters.

The member galaxies of each lensing cluster were modeled with a dual Pseudo-Isothermal Elliptical (dPIE) mass profile \citep{Eliasdottir.Limousin.ea2007} with parameters set by the above scaling relation. The introduction of core and scale radii ($r_a$, $r_s$) to a traditional isothermal profile provides a physically reasonable finite mass distribution, with $r_a$ and $r^{piv}_s$ set at 0.15 kpc and 10 kpc, respectively.

We next define the model representing the dark matter halo for each cluster. We parameterize this model with a GNFW mass density profile, a generalized adaptation of the Navarro-Frenk-White profile introduced by \citet{NFW_1996, NFW_1997}:

\begin{equation}\label{NFW}
    \rho(r) \propto \frac{1}{(r/r_s)(1+r/r_s)^2}
\end{equation}
in which the mass density $\rho(r)$ of a dark matter halo is related to scale radius $r_s$. The NFW profile has a logarithmic inner slope \citep{Schmidt2007}: \[ \gamma^{DM} = \left. -\frac{d\log\rho}{d\log r} \right|_{r \to 0} = 1 \] and provides a ``universal" mass density profile for dark matter halos in a hierarchical clustering universe. To eliminate potential bias towards the slope of $\gamma^{DM}=1$, we opt to use a GNFW profile in our modeling, where $\gamma^{DM}$ is allowed as a free variable \citep{Schmidt2007}:

\begin{equation}\label{GNFW}
    \rho(r) = \frac{\rho_s}{(r/r_s)^{\gamma^{DM}}(1+r/r_s)^{3-{\gamma^{DM}}}}
\end{equation}

Variation from the CDM prediction of $\gamma^{DM}=1$ may indicate potential physical attributes of the system, such as the presence of non-cold dark matter or baryonic feedback suggested by a shallow dark matter slope ($\gamma^{DM}<1$) \citep[e.g.][]{Spergel00,Pontzen12,Tulin18}.

Central cluster galaxies are a distinct population from other cluster members, and their kinematic properties may not follow the same scaling \cite{Bianchetti.Grillo.ea2026a}. Accordingly, in each system we model the BCG independently of the Faber-Jackson relation linking the other cluster member galaxies. The core radius is set as $r^{BCG}_a$ = 0.15 kpc and a Uniform prior is applied to the scale radius $r^{BCG}_s$ with an interval dependent on the lensing system as shown in Table \ref{table: priors}. Additionally, we include an external shear that models a warp factor that represents the cumulative gravitational influence of nearby structures on the cluster systems or complexities in the mass distribution not represented in the mass model \citep[e.g.][]{Wong11,Etherington24}. For this, we utilize \texttt{PyAutoGalaxy}'s \citep{pyautogalaxy} External Shear mass profile, which is parameterized by cartesian and diagonal components $\gamma^{ext}_1$ and $\gamma^{ext}_2$, describing xy directional shear and diagonal shear, respectively.

\subsection{Prior Configurations and Sampling}
\label{subsec:sampl}

\begingroup
    \setlength\extrarowheight{5pt}
    \setlength{\tabcolsep}{8pt} 
    \begin{table*}[!t]
        \centering
        \begin{tabular}{c | c c}
         & \shortstack{Prior\\Configuration} & Posterior \\
        \toprule \toprule
        DESJ0424 & & \\
        \midrule[\heavyrulewidth]
        $\sigma^{piv}$ (km/s) & $\mathcal{N}$(200, 20) & $177^{+15}_{-16}$ \\
        $\alpha$ (Faber-Jackson) & $\mathcal{N}$(0.27, 0.03) & $0.28^{+0.027}_{-0.026}$ \\
        Halo Mass ($log_{10}(\frac{M_{200m}}{M_{\bigodot}})$) & $\mathcal{}{U}$(14.5, 16.0) & $14.77^{+0.13}_{-0.12}$ \\
        Halo Concentration (200m) & $\mathcal{U}$(5, 12) & $8.0^{+1.4}_{-1.5}$ \\
        $\gamma^{DM}$ (GNFW) & $\mathcal{N}$(1.0, 0.2, 0.0, 2.0) & $0.85^{+0.09}_{-0.08}$ \\
        $\sigma^{BCG}$ (km/s) & $\mathcal{N}$(360, 100, 100, --) & $309^{+42}_{-39}$ \\
        $r^{BCG}_s$ (kpc) & $\mathcal{U}$(10, 50) & $14^{+11}_{-3}$ \\
        $\gamma^{ext}_1$ & $\mathcal{N}$(0.0, 0.05, -0.2, 0.2) & $0.077^{+0.010}_{-0.011}$ \\
        $\gamma^{ext}_2$ & $\mathcal{N}$(0.0, 0.05, -0.2, 0.2) & $0.030^{+0.023}_{-0.031}$ \\
        \midrule[\heavyrulewidth]
        DESJ2349 & & \\
        \midrule[\heavyrulewidth]
        $\sigma^{piv}$ (km/s) & $\mathcal{N}$(200, 20) & $196^{+16}_{-17}$ \\
        $\alpha$ (Faber-Jackson) & $\mathcal{N}$(0.27, 0.03) & $0.27^{+0.027}_{-0.029}$ \\
        Halo Mass ($log_{10}(\frac{M_{200m}}{M_{\bigodot}})$) & $\mathcal{U}$(13.5, 15.5) & $14.02^{+0.35}_{-0.28}$ \\
        Halo Concentration (200m) & $\mathcal{U}$(5, 15) & $8.6^{+4.0}_{-2.6}$ \\
        $\gamma^{DM}$ (GNFW) & $\mathcal{N}$(1.0, 0.2, 0.0, 2.0) & $0.89^{+0.16}_{-0.17}$ \\
        $\sigma^{BCG}$ (km/s) & $\mathcal{N}$(360, 100, 50, --) & $305^{+43}_{-60}$ \\
        $r^{BCG}_s$ (kpc) & $\mathcal{U}$(10, 100) & $64^{+25}_{-31}$ \\
        $\gamma^{ext}_1$ & $\mathcal{N}$(0.0, 0.05, -0.2, 0.2) & $-0.009^{+0.038}_{-0.047}$ \\
        $\gamma^{ext}_2$ & $\mathcal{N}$(0.0, 0.05, -0.2, 0.2) & $-0.014^{+0.032}_{-0.033}$ \\
        \midrule[\heavyrulewidth]
        DESJ2351 & & \\
        \midrule[\heavyrulewidth]
        $\sigma^{piv}$ (km/s) & $\mathcal{N}$(200, 20) & $197^{+22}_{-19}$ \\
        $\alpha$ (Faber-Jackson) & $\mathcal{N}$(0.27, 0.03) & $0.27\pm{0.03}$ \\
        Halo Mass ($log_{10}(\frac{M_{200m}}{M_{\bigodot}})$) & $\mathcal{U}$(13.5, 15.5) & $14.44^{+0.27}_{-0.25}$ \\
        Halo Concentration (200m) & $\mathcal{U}$(5, 15) & $10.0^{+3.2}_{-3.1}$ \\
        $\gamma^{DM}$ (GNFW) & $\mathcal{N}$(1.0, 0.2, 0.0, 2.0) & $1.05^{+0.18}_{-0.19}$ \\
        $\sigma^{BCG}$ (km/s) & $\mathcal{N}$(360, 100, 50, --) & $382^{+106}_{-103}$ \\
        $r^{BCG}_s$ (kpc) & $\mathcal{U}$(10, 100) & $62^{+28}_{-32}$ \\
        $\gamma^{ext}_1$ & $\mathcal{N}$(0.0, 0.05, -0.2, 0.2) & $0.004^{+0.036}_{-0.034}$ \\
        $\gamma^{ext}_2$ & $\mathcal{N}$(0.0, 0.05, -0.2, 0.2) & $0.005^{+0.037}_{-0.033}$ \\ 
        \end{tabular}
        \caption{Table of configured priors used for each lensing system. $\mathcal{N}$($\mu$, $\sigma$, l, u) denotes the use of a Gaussian prior of mean $\mu$ and variance $\sigma^2$, with an optional lower or upper bound (l, u). Similarly, $\mathcal{U}$(l, u) denotes a Uniform prior used within lower and upper bounds l and u. Results are reported as the posterior median with 16th and 84th percentile uncertainties.}
        \label{table: priors}
    \end{table*}
\endgroup

Our model setup and prior configuration was performed with the gravitational lensing framework provided by \texttt{PyAutoLens} \citep{pyautolens, Nightingale2015, Nightingale2018}. This software supplies analytic mass profiles, ray-tracing functions, and a model-fitting framework to compute lensing deflections and optimize model parameters. Table \ref{table: priors} lists the priors used for each sub-component of the model.

For our sampling process, we have chosen to use \texttt{dynesty}'s \citep{dynesty, Koposov2025, Skilling2004, Skilling2006, Higson2018, Feroz2009} dynamic nested sampling algorithm, which calculates both the Bayesian evidence and the Bayesian posterior distribution for the models we have defined. The nested sampling technique involves the exploration of the parameter space we have outlined for a model in which regions of low-likelihood are replaced with progressively higher-likelihood points, resulting in areas of best-fit at the conclusion of this process. We use the dynamic nested sampling method offered by \texttt{dynesty}, which allows the algorithm to adaptively allocate extra samples to regions of interest following an initial nested sampling run.

Our models are evaluated with the source-plane likelihood described in Section III.B.2 of \citet{ODonnell_2026}. Given a feature $F$ observed at a set of positions $\{\theta_i^F\}$ with uncertainties described by covariance matrices $\{\Sigma_i^F\}$, the mass model is used to predict source-plane positions and uncertainties $\{(\beta_i^F, B_i^F)\}$ for each feature. The likelihood is the probability that all positions are consistent with the best-fit for the feature $(\hat{\beta}^F, \hat{B}^F)$:

    \begin{equation}
        \log\mathcal{L}^F = -\log\sqrt{\frac{|\hat{B}^F|}{2\pi}} - \frac{1}{2}\sum_{i} (\beta_i^F - \hat{\beta}_i^F)^T (\hat{B}^F)^{-1} (\beta_i^F - \hat{\beta}_i^F)
       %log\mathcal{L} = -\frac{1}{2} \left(\chi^{2} + \mathcal{N} \right) 
    \end{equation}
    
An ideal model correctly maps each observed feature to the same source-plane position, minimizing each $\beta_i^F-\hat{\beta}^F$ and maximizing $\log \mathcal{L}^F$.

\subsection{Sampler Configuration}
\label{subsec:config}

To begin the dynamic sampling process, we specify the initial number of live-points ($n_\mathrm{live}^\mathrm{init}$) to be used in the first full nested sampling run. After this initial run, \texttt{dynesty} samples over the same parameter space again with a specified number of batch live-points ($n_\mathrm{live}^\mathrm{batch}$), and continues these batch runs as many times as needed. The sampling process is completed when the effective sample count ($n_\mathrm{eff}$) is reached.

We start the sampler with $n_\mathrm{live}^\mathrm{init}$ = 5000 points and allow \texttt{dynesty} to take additional $n_\mathrm{live}^\mathrm{batch}$ = 2000 points for each run they are needed for further expansion of the parameter space, before concluding the sampling process when $n_\mathrm{eff}$ = 300,000 is reached. For each system, we executed and combined three independent sampling runs with identical configurations to produce our final results via \texttt{dynesty}'s \texttt{merge\_runs} function, which effectively merges all samples into one run from which we can extract our results as normal. These settings were chosen after some experimentation and independent runs were used to evaluate consistency and convergence.

It is also worth noting the changes that were made to the default \texttt{dynesty} configuration parameters, Fractional Acceptance (\textit{facc}) and Posterior Fraction (\textit{pfrac}), to better accommodate the sampling behavior of our lens models. We chose to use a lower \textit{facc} value of 0.2, or 0.1 for system DESJ0424, which allows the sampler to adjust the sampler's target acceptance rate. We do keep \texttt{dynesty}'s default \textit{pfrac} value of 0.8, except for system DESJ0424, which uses a \textit{pfrac} value of 0.5. These adjustments were made to improve sampling performance for the individual models. Additionally, we use the default stopping criterion for the initial run, $\text{dlogz}_{\textrm{init}}$ = 0.01.

\subsection{X-ray Spectral Fitting}
\label{subsec:xray_fitting}

In order to determine the X-ray temperature of each cluster, we extracted and modeled the Chandra spectrum after removing background point sources. First, we created images and exposure maps in the $0.5 - 7.0$ keV energy range using \texttt{CIAO}'s \texttt{fluximage} tool. Point sources were detected in the images using \texttt{wavdetect} with a signal threshold of $2\times10^{-7}$. These point sources were then excluded from the source and background regions for the spectral fitting.

We initially defined the spectral extraction region as a 500 kpc aperture around the clusters' centers, and defined the background as an annulus region spanning from 700 to 1000 kpc, also centered on the clusters. The cluster and background spectra were extracted in an energy range of $0.5 - 7.0$ keV using the \texttt{specextract} function. For observations which had been divided based on focal-plane temperature (DESJ2349 and DESJ0424), a list of individually calibrated level-2 event files was used as the input, as opposed to a single event file. The resulting spectrum was produced by combining the individual spectra with the combined calibration files. Individual spectra were used for the separate observations of DESJ0424 and jointly fit in the modeling step described below for estimating the X-ray temperature of the cluster.

Using \texttt{CIAO}'s application for modeling and fitting X-ray data, \texttt{Sherpa} \citep{Sherpa}, we fit models to the 500 kpc spectra to derive an initial estimate of the X-ray temperature of each cluster. We used \texttt{Sherpa}'s \textit{xsapec} model to fit the emission spectrum, with an abundance of 0.3 and the known redshift of each cluster fixed, leaving the normalization and temperature to be determined by the model. \textit{xsapec} was then multiplied by \textit{xsphabs} to account for photoelectric absorption, for which we used a fixed value for the HI column density ($N_H$). We calculated $N_H$ at the RA and Dec of each lensing system using an $N_H$ calculator provided by NASA HEASARC \citep{HI4PI2016, Kalberla2005, Dickey1990}. Using \textit{xsapec}*\textit{xsphabs} results in an absorbed, optically-thin, and collisionally-ionized plasma emission model. We fit our spectra using both a spectrum binned to give 15 counts per spectral bin with the \texttt{chi2gehrels} fit statistic, and the unbinned spectra with the \texttt{wstat} statistic.  In general, we found consistent results between the two methods, and we present the final results with the \texttt{wstat} method.

Next, the cluster radius at 500 times the critical density of the universe, $r_{500c}$, was estimated based on the 500 kpc X-ray temperature using the radius-temperature scaling relation from \cite{Arnaud2005}. Cluster spectra and background spectra were then extracted for this $r_{500c}$ aperture; the background in this case was defined as an annulus spanning from 1.05$r_{500c}$ to 1.5$r_{500c}$. The $r_{500c}$ X-ray temperatures were fit using the same model as defined above. We also calculated luminosities in the 0.5-2 keV band from flux estimates (\texttt{sample\_flux}) for each cluster given by \texttt{Sherpa}'s spectral fitting. Our derived X-ray temperatures allow us to estimate the total masses, $M_{500_c}$, following the temperature-mass scaling relation defined by \cite{Mantz2016}. With fixed values of $h=0.7$ and $\Omega_m=0.3$, we adopted the values of $\beta_{0,t} = 2.16\pm{0.06}$, $\beta_{1,t}=0.63\pm{0.03}$, $\beta_{0,l} = 1.65\pm{0.14}$, and $\beta_{1,l}=1.35\pm{0.06}$ from Table 3 of \cite{Mantz2016}.

DESJ2349 proved difficult to get a good spectral fit for, as the focal-plane temperature of Chandra's CCD Imaging Spectrometer varied significantly during the observation. As a result, we elected to assume a temperature based on its known richness ($\lambda$) using the relations discussed in \cite{Kelly2024}. By fixing this temperature, we were still able to derive a luminosity measurement for DESJ2349. Since the temperature was not derived from the spectra, but rather estimated from the richness, $M_{500c}$ was instead determined using the luminosity-mass scaling relation from \cite{Mantz2016}.

% -----------------------------------------------------------------------

\section{Results}
\label{sec:result}

\subsection{Lens Modeling Results}

Following the methodology discussed in Section \ref{sec:method}, we find the lens modeling results listed in Table \ref{table: priors}. Figure \ref{fig:source_posns} in Appendix~\ref{ap:ap1} illustrates the goodness of fit by comparing the model-predicted source-plane features of our systems with their respective best-fit positions. Following the method described in Section \ref{sec:sources}, source-plane positions are estimated by our lens modeling for each of our chosen features, given the observed lens-plane positions of the features. We overlay these predicted source-plane positions with their respective best-fit source-plane positions including 1$\sigma$ elliptical uncertainties, in Figure \ref{fig:source_posns}. 

\begin{figure*}[t]
    \centering
    \subfloat[DESJ0424-3317]{%
        \includegraphics[width=0.5\textwidth]{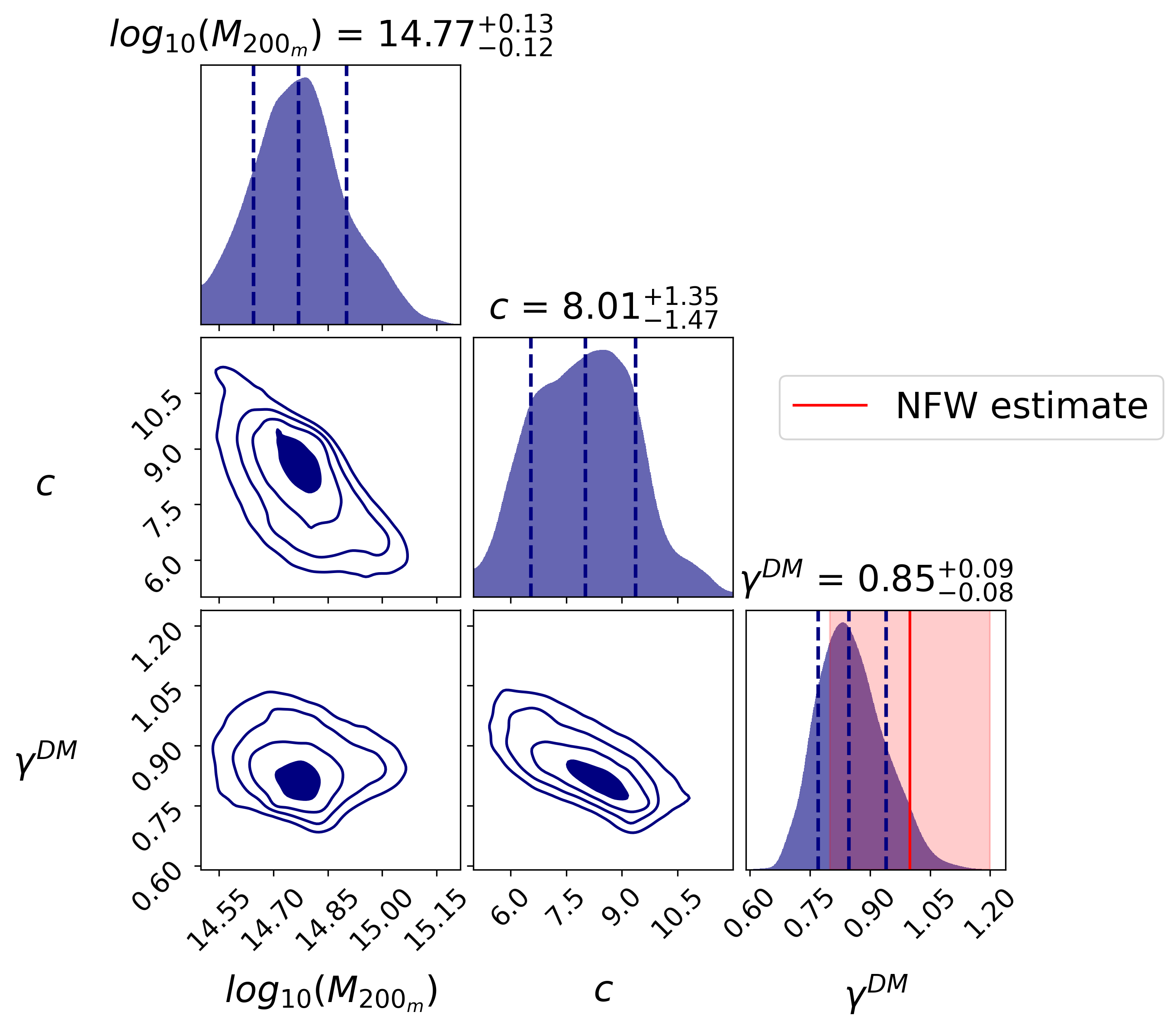}%
    }\hfill
    \subfloat[DESJ2349-5113]{%
        \includegraphics[width=0.5\textwidth]{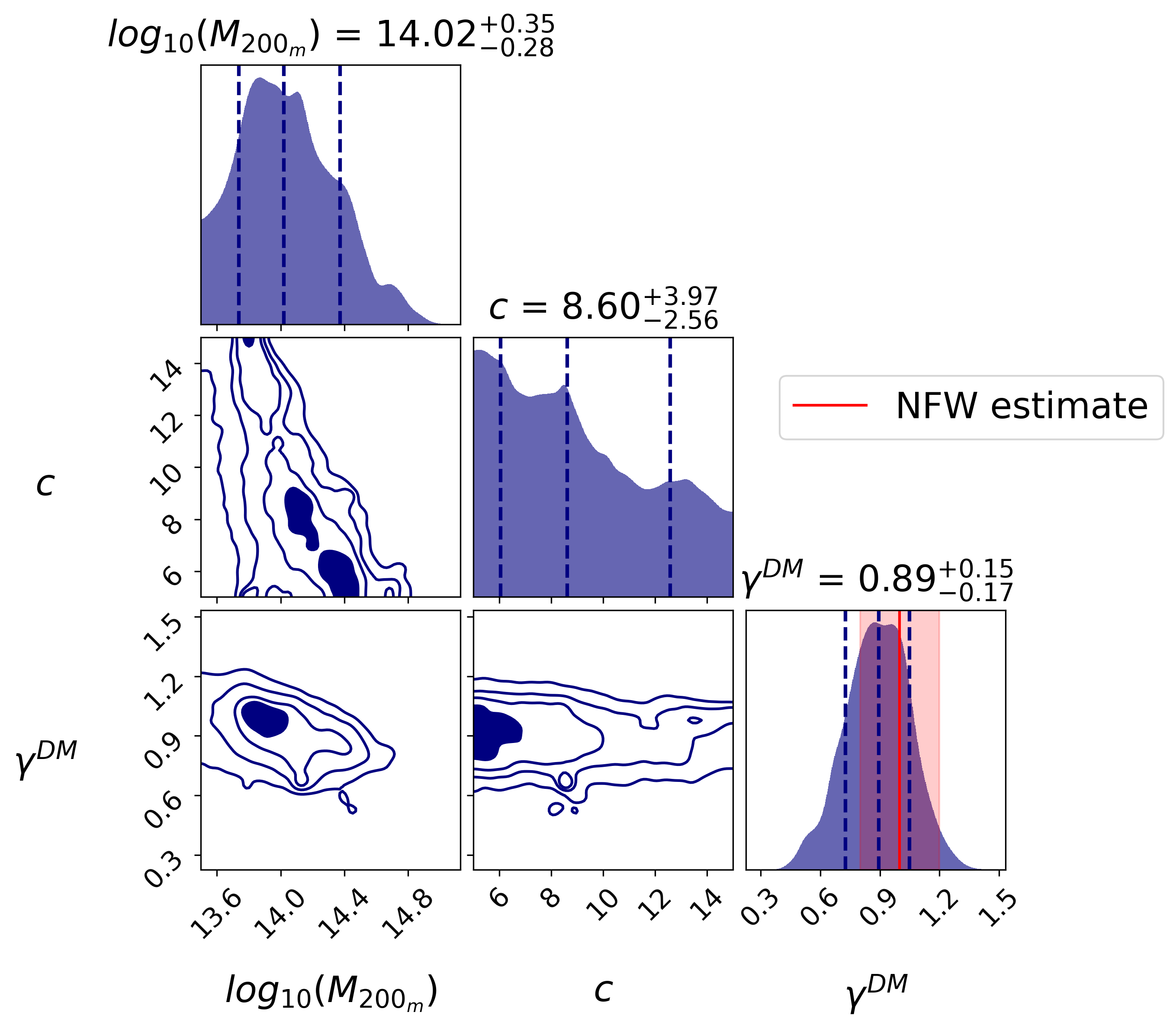}%
    }\hfill
    \subfloat[DESJ2351-5452]{%
        \includegraphics[width=0.5\textwidth]{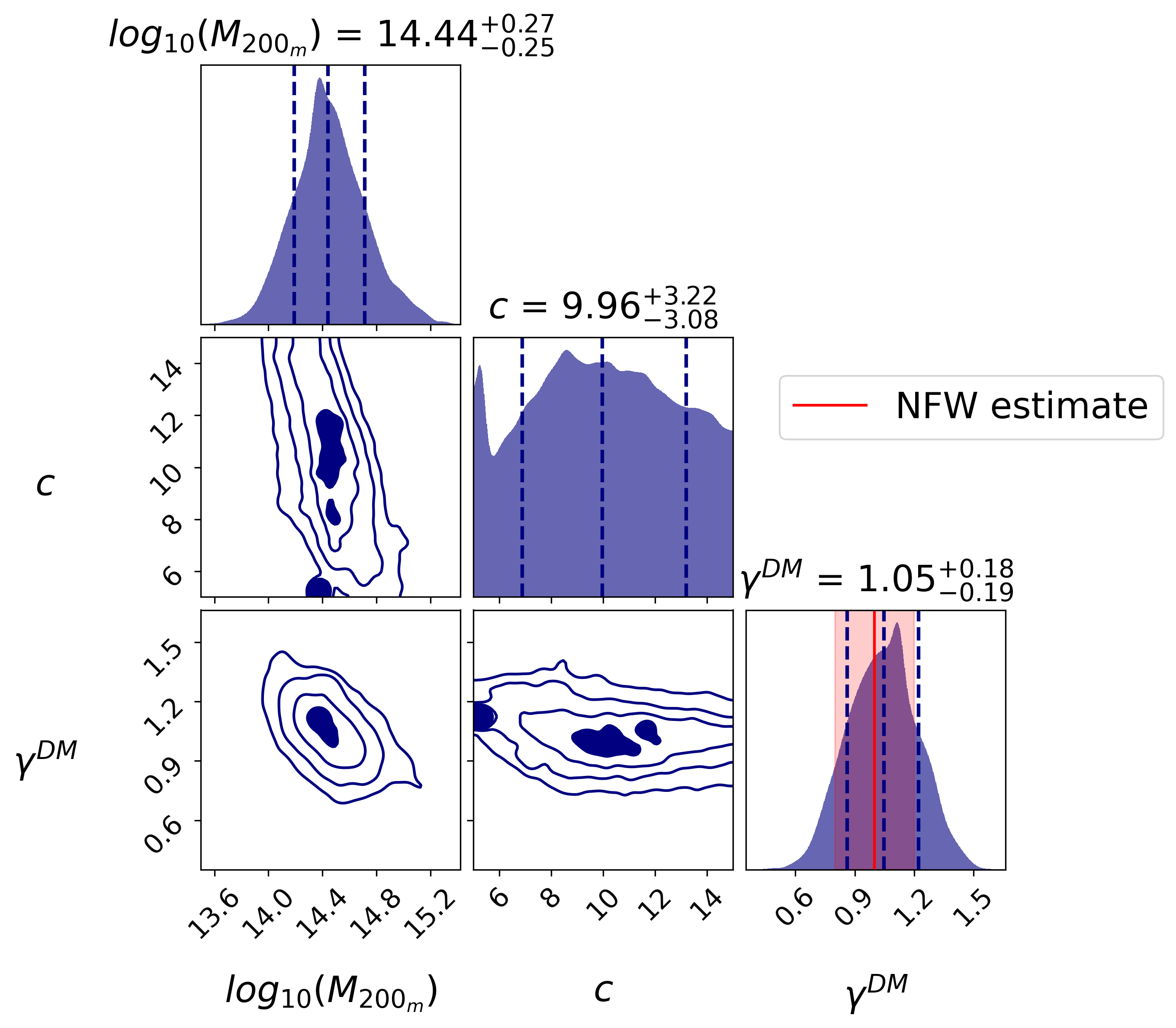}%
    }\caption{Corner plots showing the constraints on halo mass, concentration, and inner density slope. The median results and 16\% and 84\% quantiles are also shown. The red line represents both the NFW value of $\gamma=1$ and the median value of the Gaussian prior used for our GNFW slope, while the shaded region represents the 1$\sigma$ width of this prior, $\gamma=1\pm{0.2}$.}
    \label{fig:cplots}
\end{figure*}

For each fit parameter, we report the median values with $1\sigma$ uncertainties defined using the 16th and 84th percentiles of the posteriors. The parameter constraints and degeneracies are shown for the mass, concentration, and inner slope in corner plots in Figure \ref{fig:cplots}. We find logarithmic halo mass values $\log_{10}(M_{200m}/M_\odot)$ of $ \approx 14.8\pm 0.1$, $14.0^{+0.4}_{-0.3}$, and $14.4\pm0.3$ for systems DESJ0424, DESJ2349, and DESJ2351 respectively. We find dark matter concentration values of $c_{200m}^{0424} \approx 8.0^{+1.4}_{-1.5}$, $c_{200m}^{2349} \approx 9.0^{+4.0}_{-3.0}$, and $c^{2351}_{200m} \approx 10.0\pm{3.0}$, although the concentrations for DESJ2349 and DESJ2351 are poorly constrained within their prior ranges and therefore should not be considered reliable measurements. These best-fit concentrations are higher than cosmological mass-concentration relations \citep[e.g.][]{Ludlow2016} but consistent with expectations given the large uncertainties. Furthermore, we find dark matter inner slope results that are consistent between clusters: $\gamma^{DM}_{0424} \approx 0.85^{+0.09}_{-0.08}$, $\gamma^{DM}_{2349} \approx 0.89^{+0.15}_{-0.17}$, $\gamma^{DM}_{2351} \approx 1.05^{+0.18}_{-0.19}$.  While the best fits for two of the clusters are slightly shallower than one, in all cases the inner slope is consistent with the NFW expectation of $\gamma^\text{DM}=1$.
 
For the external shear parameters, we find that both components of the shear are low for all three clusters. For DESJ0424 we find $\gamma^{ext_{0424}}_1 = 0.08\pm{0.01}$ and $\gamma^{ext_{0424}}_2 = 0.03^{+0.02}_{-0.03}$. For DESJ2349, these values are $\gamma^{ext_{2349}}_1 = -0.01^{+0.04}_{-0.05}$ and $\gamma^{ext_{2349}}_2 = -0.01\pm{0.03}$, and we obtain $\gamma^{ext_{2351}}_1 = 0.004^{+0.04}_{-0.03}$ and$\gamma^{ext_{2351}}_2 = 0.005^{+0.04}_{-0.03}$ for DESJ2351. We find the overall magnitude of external shear to be $|\gamma^{ext_{0424}}| = 0.09\pm{0.01}$, $|\gamma^{ext_{2349}}| = 0.05^{+0.03}_{-0.02}$, and $|\gamma^{ext_{2351}}| = 0.04^{+0.03}_{-0.02}$. The external shear values for DESJ2349 and DESJ2351 are low and consistent with zero.  The external shear for DESJ0424 is consistent with typical values found in clusters.

\subsection{X-ray Modeling Results}
\label{sec:x-ray_temps}

The X-ray surface brightness distributions as shown in Figure~\ref{fig:x-ray} indicate that all three clusters are relatively relaxed.  The X-ray distributions, extending well beyond the strong lensing region, are generally symmetric with no indication of significant substructure.

We have measured the following X-ray temperatures from our Chandra data sets, which are consistent with expected values based on the richness of each galaxy cluster \citep{Kelly2024}: $T_X^{0424} = 2.7^{+0.7}_{-0.5}$ keV and $T_X^{2351} = 6.2^{+3.4}_{-2.0}$ keV. We also report X-ray luminosities of $L_X^{0424} = 6.0\pm{0.6} \times 10^{44}$ erg/s, $L_X^{2349} = 7.3\pm{1.7} \times 10^{43}$ erg/s, and $L_X^{2351} = 2.2\pm{0.3} \times 10^{44}$ erg/s, with 1$\sigma$ errors determined from the uncertainty on the fit parameters for both our reported X-ray temperature and luminosity results. The $M^{xray}_{500c}$ masses estimated for the clusters DESJ0424, DESJ2349, and DESJ2351, respectively, are as follows: $4.2^{+0.9}_{-0.8} \times 10^{14} M_\odot$, $1.6\pm{0.4} \times 10^{14} M_\odot$, and $4.7^{+4.1}_{-2.5} \times 10^{14} M_{\odot}$. The uncertainty on the estimated mass includes the uncertainty in the temperature and luminosity values along with the uncertainties in the scaling relation parameters given by \cite{Mantz2016} as described in Section \ref{subsec:xray_fitting}. These results are listed in Table \ref{table:cxo}.

\begingroup 
    \renewcommand{\arraystretch}{1.5} 
    \setlength\extrarowheight{2pt}
    \setlength{\tabcolsep}{8pt} 
    \begin{table*}[t]
        \centering
        \begin{tabular}{c || c c c c c}  
            System & $T_X$ (keV) & $L_X$ (erg/s) & $\lambda$ & $M^{xray}_{500c}$ (10$^{14} M_\odot$) & $M^{lens}_{500c}$ ($10^{14} M_\odot$) \\
            \hline     
            DESJ0424-3317 & $2.7^{+0.7}_{-0.5}$ & $6.0\pm{0.6} \times 10^{44}$ & 54 & $4.2^{+0.9}_{-0.8}$ & $3.9^{+1.4}_{-0.9}$ \\
            DESJ2349-5113 & $4.5^\dag$ & $7.3\pm{1.7} \times 10^{43}$ & 65 & $1.6\pm{0.4}$ & $0.7^{+0.8}_{-0.3}$ \\
            DESJ2351-5452 & $6.2^{+3.4}_{-2.0}$ & $2.2\pm{0.3} \times 10^{44}$ & 83 & $4.7^{+4.1}_{-2.5}$ & $1.8^{+1.6}_{-0.8}$ \\
        \end{tabular}
        \caption{Table of derived x-ray temperatures, luminosities, and masses for each cluster, along with their redMaPPer richness $\lambda$. For comparison, this table also reports our lens-derived masses converted to $M_{500c}$.\\
        $^\dag$DESJ2349 had large focal-plane temperature variations during the observation and we were unable to robustly measure the temperature. We use a fixed 4.5 keV based on the known richness to estimate the luminosity and the mass is estimated from $L_X$.}
        \label{table:cxo}
    \end{table*}
\endgroup

% -----------------------------------------------------------------------

\section{Discussion}
\label{sec:disc}

We chose to model the cluster mass distributions with a generalized-NFW mass density profile to test for potential flattening of the inner slope that might result from non-CDM dark matter scenarios. However, we find no indication of deviations of the inner slope from the NFW expectation. Our resulting inner slope values are well constrained, ranging from approximately $0.85\pm{0.1}$ to $1.05\pm{0.2}$, which are consistent with a slope of $\gamma^{DM}_{NFW}=1$. This result is in contrast to several previous works which found shallow inner slopes or a preference for a core \citep{Sand04,Newman13b,Limousin22,Cenry25}, though other works find inner slopes consistent with NFW \citep{Sartoris20,Biviano23}. Notably, the recent work of \citet{Cenry25} finds an average inner slope of $\sim\!0.66$ for a sample of six strong lensing clusters, but these clusters are selected to have radial arcs, which preferentially appear in clusters with shallower density profiles.

As depicted in Figure \ref{fig:cplots}, our median dark matter concentration results are fairly consistent between clusters within our data set, but with large errors indicating poor constraints on this parameter. Our $c_{200m}$ concentration values, approximately ranging from $8^{+1.4}_{-1.5}$ to $10\pm{3}$, are consistent with the lensing-derived result of $c_{200m} \approx 10^{+2}_{-3}$ for the MACS J0138-2155 galaxy cluster modeled by \cite{ODonnell_2026}. Although the best-fit concentration values are higher than expected for galaxy clusters in general, the uncertainties are large and strong lensing clusters tend to exhibit these higher concentrations \citep[e.g.][]{Comerford2007}, particularly relaxed strong lensing clusters \citep{Giocoli2014}. 

Two of our three clusters (DESJ2349 and DESJ2351) are well described by a single, spherical halo model with external shear consistent with zero, and only DESJ0424 has a non-negligible $\gamma^{ext}$. This suggests that the strong-lensing models of these clusters do not require additional substructure or ellipticity beyond that of a single, approximately symmetric halo. This is notable given that substantial external-shear terms ($\gamma^{ext} \sim\!0.07-0.17$) have been required in strong-lensing models of several clusters \citep[e.g.][]{Mahler18,Caminha19,Lagattuta19,Bergamini19,Acebron22,ODonnell_2026}. At galaxy scales, recent studies likewise find substantial fitted shear to be common \citep[e.g.][]{Hogg26}. While sometimes discussed as shear due to line of sight structure, the external shear values are high for this interpretation, and \citet{Etherington24} show that external shear is generally due to a lack of complexity in the lens model. Likewise, \citet{Limousin25} show for the case of the Abell 370 cluster that they can remove the need for an external shear term by adding additional cluster substructure. The negligible shear in our systems therefore indicates that they have unusually simple projected mass distributions and is consistent with their being relatively relaxed systems. For DESJ0424, the magnitude of the external shear ($|\gamma^{ext_{0424}}|$) is $0.09\pm{0.01}$; for this cluster, both the arc structure and X-ray show mild ellipticity, perhaps indicating the need for some elongation in the lens model.  The structure of the source galaxy in this system is also particularly complex and could potentially benefit from a pixel-level modeling.

We can compare our independently estimated cluster masses based on our lensing results to those estimated from the X-ray temperatures. In order to compare to $M^{xray}$, which have been estimated as $M_{500c}$ masses, the lensing-derived masses $M^{lens}$ have been converted from $M_{200m}$ to $M_{500c}$ using the \texttt{Colossus} \citep{Diemer_2018} toolkit, assuming Planck18 cosmology \citep{Planck18} and given the $z_{lens}$ of each cluster. The adjusted lensing masses are given in Table \ref{table:cxo}. 
The two methods of estimating the cluster masses generally agree well. The differences in mass between  approaches are within 1$\sigma$ in all cases.

The slope of the Faber-Jackson relation $\alpha$ is nearly unconstrained by our data, with all three systems very nearly reproducing the prior of $\mathcal{N}(\mu=0.3, \sigma=0.03)$. Since $\alpha$ exhibits very little degeneracy with other model parameters, its value is unlikely to affect our other results.

% -----------------------------------------------------------------------

\section{Conclusion}
\label{sec:conc}

We used a combination of strong lens modeling and Chandra X-ray observations to model the structure and dark matter content of the three strong gravitational-lensing galaxy clusters, DESJ0424-3317, DESJ2349-5113, and DESJ2351-5452, found in the Dark Energy Survey \citep{ODonnell2022}. Both the symmetric lensing giant arc structures and, on larger scales, the X-ray surface brightness distributions indicate that these are relaxed clusters ideal for strong lens modeling. In fact, two of the three clusters are well modeled by a single, spherical dark matter halo plus contributions from the BCG and member galaxies with low external shear, which might have indicated the need for additional complexity in the lens model.
 
We find cluster lensing estimated masses between $10^{14}$ and $6 \times 10^{14}$ $M_\odot$. In all cases, the masses estimated from the X-ray temperature or luminosity are consistent with the lensing mass, and both the X-ray properties ($T_X, L_X$) and masses are consistent with the DES cluster richness. 

In our lens models, the main dark matter halo is parameterized with a GNFW mass density profile allowing for a variable inner slope. While the best-fit inner slope is somewhat shallower than 1 for two of the clusters, we find inner slope consistent with 1 in all three clusters, as expected for an NFW profile. Thus we find no indication of a cored dark matter profile, as would be expected in non-CDM dark matter scenarios such as SIDM.  Future Integral Field Unit (IFU) spectroscopy would allow us to more precisely constrain the inner density profile through the combination of resolved kinematics of the BCG with lensing, as was demonstrated in \cite{ODonnell_2026}. The relaxed nature and large, symmetric giant arc systems make these ideal targets for constraints on SIDM.
\vspace{0.2cm}

% -----------------------------------------------------------------------

\section*{Acknowledgments}

Support for this work was provided by the National Aeronautics and Space Administration through Chandra Award Number GO3-24103X issued by the Chandra X-ray Center, which is operated by the Smithsonian Astrophysical Observatory for and on behalf of the National Aeronautics Space Administration under contract NAS8-03060. This research is based on observations made with the NASA/ESA Hubble Space Telescope obtained from the Space Telescope Science Institute, which is operated by the Association of Universities for Research in Astronomy, Inc., under NASA contract NAS 5–26555. These observations are associated with programs 16773 and 11974.
Additionally, we acknowledge that this research has made use of data obtained from the Chandra Data Archive provided by the Chandra X-ray Center (CXC).

In addition to the several open-source Python packages previously referenced throughout this work (\texttt{PyAutoLens}, \texttt{PyAutoGalaxy}, \texttt{Colossus}, \texttt{skimage}, \texttt{dynesty}, \texttt{CIAO}, and \texttt{Sherpa}), this research has also made use of additional Python packages including: \texttt{PyAutoFit} \citep{pyautofit}, \texttt{astropy} \citep{astropy1, astropy2, astropy3}, \texttt{NumPy} \citep{Harris2020}, and \texttt{matplotlib} \citep{Hunter2007}.

Additionally, we acknowledge the use of the lux supercomputer at the University of California, Santa Cruz, funded by NSF MRI grant AST 1828315.

% -----------------------------------------------------------------------

\bibliographystyle{mnras}

% You should give the same name for your .bbl as your main .tex
\bibliography{oja_template}

% -----------------------------------------------------------------------

\begin{appendix}

\section{Appendix 1}
\label{ap:ap1}
% set fig counter to 0 and set type of counter to A#
\renewcommand{\thefigure}{A\arabic{figure}}
\setcounter{figure}{0}

\begin{figure*}
    \centering
    \subfloat[DESJ0424-3317]{%
        \includegraphics[width=0.2\textwidth]{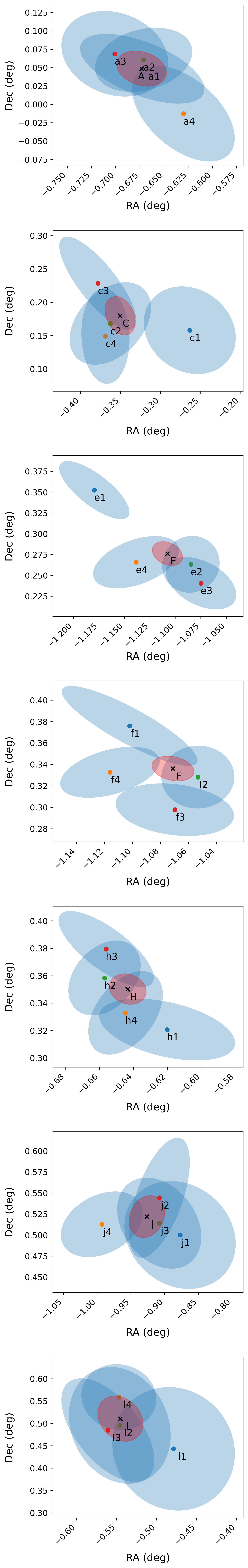}%
    }
    \subfloat[DESJ2349-5113]{%
        \includegraphics[width=0.2\textwidth]{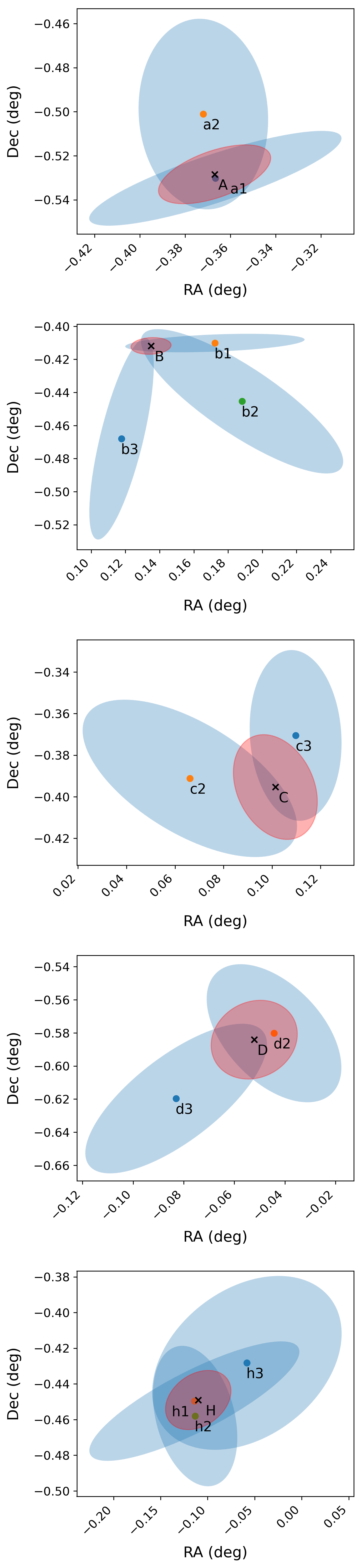}%
    }
    \subfloat[DESJ2351-5452]{%
        \includegraphics[width=0.2\textwidth]{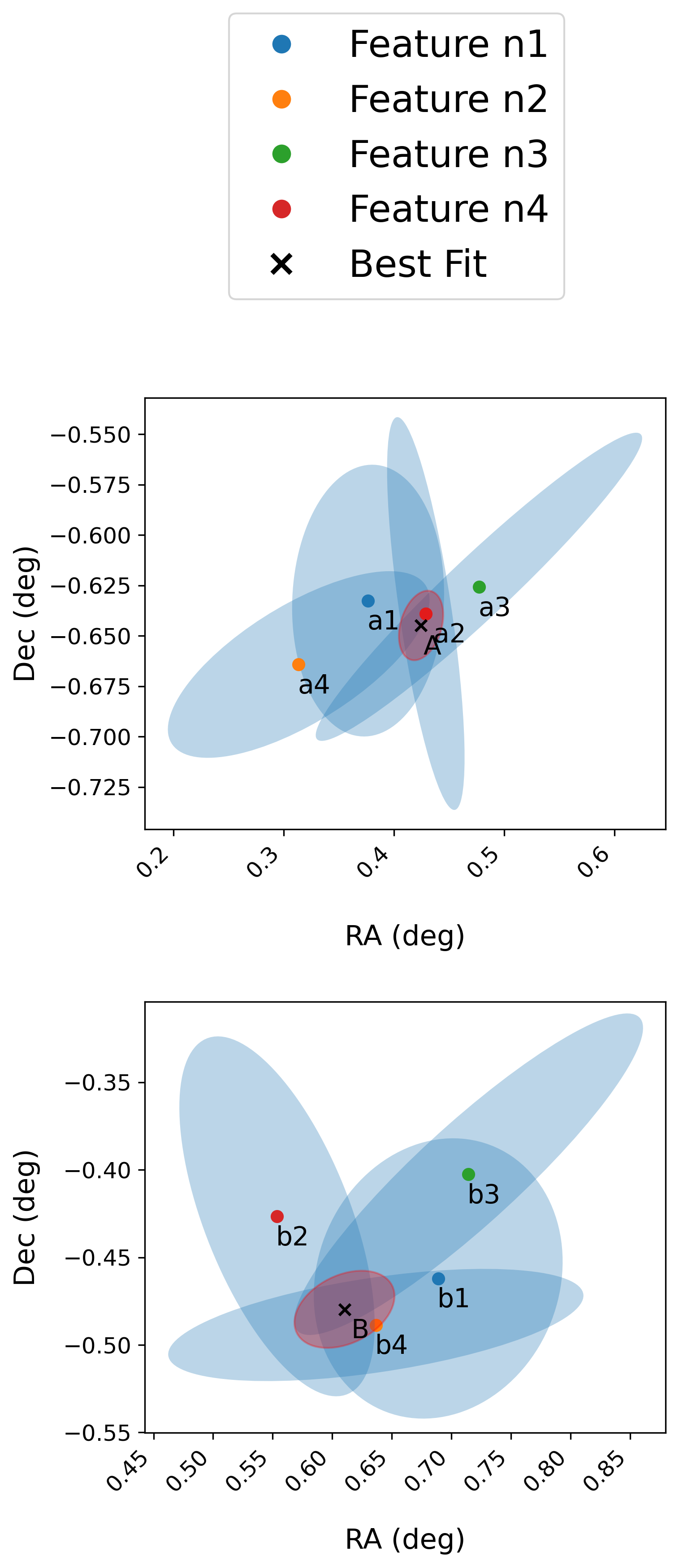}%
    }
    \caption{Reconstructed source-plane positions of each of the features used in lens modeling (e.g.~``a1"), of each system, as compared to the best-fit source-plane coordinate of that feature (e.g.~``A"). Shaded regions represent the 1$\sigma$ uncertainties on the source-plane feature positions and the best-fit coordinates.}
    \label{fig:source_posns}
\end{figure*}

Here we provide additional figures to further support our results. Figure \ref{fig:source_posns} shows our selected lensed features at their estimated source-plane positions as compared to a best-fit position for each grouping. The visual overlap of a feature's elliptical uncertainty (blue) with the best-fit ellipse (red) allows for an assessment of the goodness of fit of our models to these systems. 

Figures \ref{fig:cplot_all_0424}, \ref{fig:cplot_all_2349}, and \ref{fig:cplot_all_2351} expand upon Figure \ref{fig:cplots}, showing the results of our full list of variables used.

\begin{figure*}
    \centering
    \includegraphics[width=1\textwidth]{figs/0424_cPlot_all.png}
    \caption{Corner plot for system DESJ0424 showing the constraints on all parameters. The posterior medians and 16\% \& 84\% quantiles are shown.}
    \label{fig:cplot_all_0424}
\end{figure*}
\begin{figure*}
    \centering
    \includegraphics[width=1\textwidth]{figs/2349_cPlot_all.png}
    \caption{Corner plot for system DESJ2349 showing the constraints on all parameters. The posterior medians and 16\% \& 84\% quantiles are shown.}
    \label{fig:cplot_all_2349}
\end{figure*}
\begin{figure*}
    \centering
    \includegraphics[width=1\textwidth]{figs/2351_cPlot_all.png}
    \caption{Corner plot for system DESJ2351 showing the constraints on all parameters. The posterior medians and 16\% \& 84\% quantiles are shown.}
    \label{fig:cplot_all_2351}
\end{figure*}

\end{appendix}

\end{document}